\documentclass[aps,prd,nofootinbib,superscriptaddress,preprint,eqsecnum,showkeys,showpacs,preprintnumbers]{revtex4-2}
\usepackage{amsmath}
\usepackage{amssymb}
\usepackage{graphicx}
\usepackage{epsfig}
\usepackage{xcolor,color}
\usepackage{url}
\usepackage{bm}
\usepackage{mathrsfs}
\usepackage[utf8]{inputenc}
\usepackage{hyperref}
\usepackage{enumerate}
\usepackage{amsthm}
\usepackage{bbm}
\usepackage[normalem]{ulem}
\usepackage{upgreek}
\usepackage{tensor}
\usepackage{siunitx}
\usepackage{orcidlink}
\usepackage[caption=false]{subfig}
\usepackage{colortbl}
\usepackage{hyperref} 
\usepackage{appendix} 
\usepackage{cleveref}
\usepackage{multirow}
\usepackage{booktabs}
\usepackage{threeparttable}
\usepackage[export]{adjustbox}
\usepackage{times}
\usepackage{commath}
 
 \allowdisplaybreaks[3]
\hypersetup{colorlinks=true,linkcolor=blue,anchorcolor=blue,citecolor=blue}  
\definecolor{colour1}{HTML}{0571b0} 
\definecolor{colour2}{HTML}{92c5de} 
\definecolor{colour3}{HTML}{f4a582} 
\definecolor{colour4}{HTML}{ca0020} 
\definecolor{colour5}{HTML}{fe4a49} 
\definecolor{colour6}{HTML}{2d3092} 

\hypersetup{colorlinks=true, linkcolor=colour6, citecolor=colour6,
filecolor=colour6, urlcolor=colour6}

\theoremstyle{plain}

\newcommand{\bea}{\begin{eqnarray*}}
\newcommand{\eea}{\end{eqnarray*}}
\newcommand{\bean}{\begin{eqnarray}}
\newcommand{\eean}{\end{eqnarray}}

\newcommand{\vvr}{\mbox{\boldmath${r}$}}
\newcommand{\vp}{\mbox{\boldmath${p}$}}
\newcommand{\vP}{\mbox{\boldmath${P}$}}

\newcommand{\vS}{\mbox{\boldmath${S}$}}

\newcommand{\ud}{\mathrm{d}}

\newcommand{\ph}[1]{\phantom{#1}}

\newcommand{\be}{\begin{equation}}
\newcommand{\ee}{\end{equation}}

\newcommand\beq{\begin{equation}}
\newcommand\eeq{\end{equation}}
\def\bea{\begin{eqnarray}}
\def\eea{\end{eqnarray}}

\begin{document}

\title{Evolution dynamics of spinning binaries  in effective-one-body theory to fifth Post-Minkowskian order }

\author{Jiliang Jing\orcidlink{0000-0002-2803-7900}\footnote{jljing@hunnu.edu.cn}}
 \affiliation{Department of Physics, Key Laboratory of Low Dimensional Quantum Structures and Quantum Control of Ministry of Education, and Synergetic Innovation
Center for Quantum Effects and Applications, Hunan Normal
University, Changsha, Hunan 410081, P. R. China}
\affiliation{Center for Gravitation and Cosmology, College of Physical Science and Technology, Yangzhou University, Yangzhou 225009, P. R. China}

\author{Sheng Long
}
\affiliation{School of Fundamental Physics and Mathematical Sciences, Hangzhou Institute for Advanced Study, University of Chinese Academy of Sciences, Hangzhou 310024, China}

\author{Weike Deng\orcidlink{0009-0002-5504-8151}
} 
\affiliation{School of Science, Hunan Institute of Technology, Hengyang 421002, P. R. China}

\author{Jieci Wang\footnote{ jcwang@hunnu.edu.cn}} 
\affiliation{Department
of Physics, Key Laboratory of Low Dimensional Quantum Structures and
Quantum Control of Ministry of Education, and Synergetic Innovation
Center for Quantum Effects and Applications, Hunan Normal
University, Changsha, Hunan 410081, P. R. China}

\begin{abstract}

This paper investigates a self-consistent effective-one-body (EOB) theory developed to describe the dynamics of spinning binary systems. We systematically derive, using field-theoretic principles, all expressions and quantities entering Hamilton's equations, which govern the dynamical evolution of the system. Based on the effective rotating metric up to fifth post-Minkowskian (PM) order recently obtained by us, we construct the Hamiltonian and the stress-energy tensor. We then derive a decoupled, separable Teukolsky-like equation with a source term for the Newman--Penrose Weyl scalar $\psi_4^B$ and present its formal solution. Finally, using this solution, we obtain the energy flux, radiation-reaction force, and gravitational waveforms for the ``plus'' and ``cross'' polarizations generated by spinning binaries. The EOB theory applies to systems with arbitrary mass ratio, which  filling the gap between comparable mass ratios and extreme mass ratios.

\end{abstract}


\maketitle

\newpage


 \section{Introduction}
 
 Gravitational waves emitted during the inspiral, plunge, and coalescence phases of binary systems, whether nonspinning or spinning, consisting of a stellar-mass black hole (or a compact object of comparable mass) and a massive black hole with mass ranging from $M \sim 10^4$ to $10^7 M_\odot$, are expected to be detectable in the frequency bands of LISA, the European Space Agency's L3 mission, as well as Taiji and TianQin. Extreme mass-ratio inspirals (EMRIs) enable precise measurements of the properties of massive black holes and their stellar environments \cite{VigeHugh10, BaraCutl07, BrowETC07, Gair:2012nm, Barausse:2014tra}. Another important class of systems involves intermediate-mass black holes (IMBHs) with masses of approximately $10^2$ to $10^3 M_\odot$. When a stellar-mass compact object inspirals into an IMBH, the resulting binary may merge within the detection band of ground-based observatories if its chirp mass is $\lesssim 350 M_\odot$ \cite{BrowETC07}. Moreover, the inspiral phase of such systems may be observable by LISA weeks before merger in the LIGO 
 band \cite{Miller:2002vg, Sesana:2016ljz}. Conversely, when an IMBH inspirals into
a massive black hole, the resulting intermediate mass-ratio inspiral can produce a particularly strong signal in the LISA band.
  
In the detection of gravitational waves (GWs) ~\cite{Abbott2016, Abbott20162, Abbott2017, Abbott20172, Abbott20173, Abbott2019, Abbott20211, Abbott20212, Abbott20213}, the dynamical evolution of binary systems is essential for accurately estimating the physical parameters associated with the inspiral and coalescence of relativistic compact binaries. A novel method for investigating these dynamics is the EOB theory, which is based on the post-Newtonian (PN) approximation initially introduced by Buonanno and Damour \cite{Damour1999, Damour20002, Damour2007, Damour20091, Damour2000, Damour2001, DamourH, Pan2, Damour2009}. This theory serves as the foundation for the computation of numerous gravitational waveform  templates \cite{Taracchini, Bohe, Abbott2016, Abbott20162, Abbott2017, Abbott20172, 7g4q-5crv, PhysRevD.104.024050, PhysRevD.84.044014, PhysRevD.96.044028, PhysRevD.111.024004, ChenZhang85401, zhong2026self}.  To overcome the limitations associated with the low-velocity assumption inherent in the PN approximation and improve the model's accuracy, Damour \cite{Damour2016} proposed an alternative EOB framework based on the PM approximation, which has attracted significant attention \cite{Damour2017, Damour2018, Damour2018new, Antonelli2019, Damour2019, Damour2020, HeLin2016, Jing2019, Blanchet2018, Cheung2018, Vines2019, Cristofoli2019, Collado2019, Bern2019, Bern20192, Plefka2019, BiniDamour2020, Cheung2020, WOS:001145620900001, WOS:000696849100003, WOS:000616135600002}.
  
In a self-consistent EOB theory, the Hamiltonian, stress-energy tensor, gravitational-wave energy flux, radiation-reaction force (RRF), and waveform should all be derived from a unified physical model. In our previous work \cite{Jing}, we established a self-consistent EOB theory for the spinless two-body system based on the PM approximation, applicable to the dynamics of nonspinning black holes characterized by two mass parameters, $(m_1, m_2)$. However, it is well known that astrophysical black holes are rotating. Consequently, the dynamical evolution of spinning black-hole binaries depends on both mass and spin parameters, namely the masses ($m_1,  m_2$) and spins ($\boldsymbol{S}_1, \boldsymbol{S}_2)$, yielding a total of eight parameters. This additional complexity makes the dynamics of spinning binaries significantly richer. Therefore, a crucial next step is to extend the investigation from the spinless two-body system \cite{Jing} to binary systems composed of spinning black holes. We made initial attempts in Ref. \cite{JingXZ}. However, in that study the source of gravitational radiation did not incorporate the spin contributions of the effective test particles, and the variables associated with $\psi^B_4$ could not be directly separated. To overcome these shortcomings, we \cite{Jing2025SC} recently constructed a self-consistent EOB theory for spinless binary systems by adopting a new gauge in which the decoupled equation can be separated into variables in the general case.

In this paper, we aim to develop a self-consistent EOB theory for binary systems of spinning black holes based on the studies in Ref. \cite{Jing2025SC}. Using the effective rotating metric up to fifth PM order obtained in our previous work \cite{JingNew2025}, we derive the general formulas for the Hamiltonian and stress-energy tensor from an action principle. By incorporating this effective rotating metric into the Hamiltonian formalism, we obtain a general explicit expression for the effective Hamiltonian of the EOB theory applicable to realistic spinning two-body systems. Finally, we analyze the energy flux, the RRF, and the waveforms of the ``plus" and ``cross" modes of gravitational waves generated by spinning binary systems. These analyses are based on the decoupled and separable Teukolsky-like equation for the null-tetrad component of the perturbed Weyl tensor, $\psi^B_4$, which includes a source term determined by the stress-energy tensor in the effective spacetime.

The remainder of this paper is organized as follows. In Section II, we derive the Hamiltonian and stress-energy tensor within the EOB framework based on the effective rotating metric at fifth PM order. Section III presents a detailed formulation of the improved Hamiltonian for the spinning two-body system. Section IV addresses the formal solution of the Teukolsky-like equation, with the source term characterized by the stress-energy tensor in the effective rotating spacetime. In Section V, we present the energy flux, the RRF, and the waveforms of the plus and cross polarizations of GWs. Finally, we conclude with a summary and discussion in the last section.

Notation: In this paper, we use three types of tetrads: an orthonormal tetrad $e_{a}^{\mu}$ attached to the worldline of the effective test particle, an orthogonal tetrad $\tilde{e}^\mu_{A}$ associated with the effective background spacetime, and a null tetrad $Z_{a\mu}=(l_\mu,\, n_\mu,\, m_\mu,\, \bar{m}_\mu)$ associated with the effective background spacetime. In sections \ref{PModel} and \ref{Hamiltonian}, spacetime tensor indices (ranging from 0 to 3) are denoted by Greek letters, while lowercase Latin letters denote spatial indices (1-3). The orthonormal tetrads $e_{a}^{\mu}$ and $\tilde{e}^\mu_{A}$ satisfy $e^\mu_{ a}\,e^\nu_{ b}\, g_{\mu\nu} = \eta_{ a b}$ and $\tilde{e}^\mu_{ A}\,\tilde{e}^\nu_{ B}\, g_{\mu\nu} = \eta_{ A B}$, respectively, with $\eta_{TT}=1$, $\eta_{TI}=0$, and $\eta_{IJ}=-\delta_{IJ}$, where uppercase Latin letters ${I}$ and ${J}$ run from 1 to 3, and $T$ denotes the timelike tetrad index. The two orthonormal tetrads are related by Lorentz matrices $\Lambda_{A}^{a}$ according to
$	e_{a}^{\mu}(\phi,x) = \Lambda^{A}_{a} (\phi)\tilde{e}_{A}^{\mu}(x) \,, \ \ 
	\Lambda_{A}^{a} \Lambda_{Ba} = \eta_{AB}$ and $ \Lambda_{Aa} \Lambda^{A}_{b} = \eta_{ab} \,, 
$
and their indices satisfy $a,\,b,\,A\,,B=0\,, 1\,, 2\,, 3$. All orthonormal tetrad indices are raised and lowered with the Minkowski metric $\boldsymbol{\eta}=diag(1,-1,-1,-1)$. In section \ref{RRFWF}, the spacetime tensor indices (ranging from 0 to 3) are again denoted by Greek letters, while the null tetrad $Z_{a\mu}$ satisfies $g^{\mu\nu}Z_{a\mu} Z_{b\nu}=\tilde{\eta}_{ab}$ with the flat-space metric
$
\tilde{\eta}^{pq}=\tilde{\eta}_{pq}=\left[ 
\begin{array}{cc}
A \   & 0   \\ 
0\   & -A   
\end{array}\nonumber 
\right ],
$
where
$
A=\left[ 
\begin{array}{cc}
0 \ \ \ & 1\   \\ 
1\ \ \ & 0  
\end{array}\nonumber 
\right ]. 
$

\section{ Physical model}\label{PModel}

To ensure that all quantities appearing in the Hamilton equations governing the dynamical evolution of this system are derived from a unified physical model, we formulate the effective Hamiltonian and the stress-energy tensor associated with gravitational radiation from a single action in the effective spacetime.

\subsection{Effective rotating metric  with radiation-reaction effects  to 5PM order}

Due to the third-generation gravitational-wave detectors requiring at least fifth PM order precision \cite{Driesse_2025}, based on Bern et al.'s expression for the conservative Hamiltonian \cite{Bern20192}  of a relativistic, massive, spinless two-body system, we found that the effective metric of spinless  binaries to fifth PM order can be written as \cite{JingNew2025}
$
ds_{\text{eff}}^2=\frac{\Delta_0}{r^2} dt^2-\frac{r^2}{\Delta_0}dr^2- r^2(d\theta^2+\sin^2\theta d\varphi^2),\label{Mmetric}
$
with
$ \Delta_0=r^2- 2 GM r+\sum_{i=2}^\infty a_i \frac{(GM)^i}{r^{i-2}}, 
$
where $a_i$ (i=2, 3, 4, 5) are given in Ref. \cite{JingNew2025}.

By applying the method of constructing an effective rotating metric from the effective metric of spinless binaries, as discussed by Damour and Barausse et al. \cite{Barausse,Damour}, we further showed that, based on the effective metric for spinless binaries, the effective rotating metric to 5PM order is given by 
\cite{JingNew2025}
\begin{align}\label{effmetric}
 ds^{2}&=g^{\text{eff}}_{\mu\nu}d x^\mu d x^\nu\nonumber\\&=\frac{\Delta -a^2\sin ^2\theta}{\Sigma}dt^{2}-\frac{\Sigma}{ \Delta } dr^{2}-\Sigma d\theta^{2}-\frac{\Lambda \sin^2\theta }{\Sigma}  d\phi^{2}\nonumber \\ &+\frac{2\omega_{pm}  \sin^2\theta}{ \Sigma}dt d\varphi, 
\end{align}
where $ \Sigma=\bar{\rho} \bar{\rho}^{*},  $ $ \bar{\rho}=r+i a \cos\theta, $ $ \bar{\rho}^{*}=r-i a \cos\theta, $ $ \Delta =\Delta_0 +a^2, $ $ \Lambda=\varpi^{4}-a^{2} \Delta  \sin^{2}\theta, $ $ \varpi=(r^{2}+a^{2})^{\frac{1}{2}},$ $ \omega_{pm}=a(a^{2}+r^{2}-\Delta ),$
and $a=\left|\vS_{metric}\right|/M$ is the rotational parameter.

\subsection{Hamilton equations for EOB theory of spinning binaries }

For a spinning two-body system, the basic idea of EOB theory is to map the dynamics of two compact objects with masses ($m_1$, $m_2$) and spins ($\vS_1$, $\vS_2$) onto the dynamics of an effective test particle with mass $m_0=m_1m_2/(m_1+m_2)$ and spin $\vS=\vS(\vS_1,\vS_2)$ orbiting a massive black hole characterized by an effective rotating metric $g^{\text{eff}}_{\mu\nu}$ with mass parameter $M =m_1+m_2$ and rotational parameter $a=\left|\vS_{metric}\right|/M$, where $\vS_{metric}=\vS_{metric}(\vS_1,\vS_2)$. The EOB dynamics are governed by the Hamilton equations \cite{Damour2001,Taracchini1}
\begin{subequations} \label{HEq} 
\begin{align}  &  \frac{d\vvr}{d\hat{t}}=\{\vvr,\hat{H}[g^{\text{eff}}_{\mu\nu}]\}=\frac{\partial \hat{H}[g^{\text{eff}}_{\mu\nu}]}{\partial \vp}\,, \label{EOM2}\\ 
 &   \frac{d\vp}{d\hat{t}}=\{\vp,\hat{H}[g^{\text{eff}}_{\mu\nu}]\}+\hat{\bm{\mathcal{F}}}[g^{\text{eff}}_{\mu\nu}]=-\frac{\partial \hat{H}[g^{\text{eff}}_{\mu\nu}]}{\partial \vvr}
    +\hat{\bm{\mathcal{F}}}[g^{\text{eff}}_{\mu\nu}]\,, \label{EOM3} \\ 
&   \frac{d\vS_{1,2}}{d\hat{t}} = \{\vS_{1,2}, \nu \hat{H}[g^{\text{eff}}_{\mu\nu}] \} = \nu \frac{\partial \hat{H}[g^{\text{eff}}_{\mu\nu}]}{\partial \vS_{1,2}} \times \vS_{1,2}  +\hat{\bm{\mathcal{F}}_{1,2}^s}\,, \label{EOM4}
\end{align}
\end{subequations}
where $\hat{t}\equiv t/ M $,  $\nu=\frac{m_0}{ M }=\frac{m_1 \, m_2}{(m_1+m_2)^2}$,
$\hat{H}[g^{\text{eff}}_{\mu\nu}]=h_[g^{\text{eff}}_{\mu\nu}]/m_0$ is the reduced EOB Hamiltonian \cite{BarausseH,BarausseH1,Barausse},   $\hat{\bm{\mathcal{F}}}[g^{\text{eff}}_{\mu\nu}]=\bm{\mathcal{F}}[g^{\text{eff}}_{\mu\nu}]/m_0$ is the reduced RRF, and $\hat{\bm{\mathcal{F}}_{1,2}^s}=\bm{\mathcal{F}_{1,2}^s}/m_0$ is the spin RRF  which can be found in \cite{PhysRevD.96.084064,Liu}.

By employing the energy relation \cite{Jing,Jing1,JingXZ}, the improved reduced EOB Hamiltonian appearing in Eq. (\ref{HEq}) can be written as
$
\hat{H}[g^{\text{eff}}_{\mu\nu}]=\frac{1}{\nu} \sqrt{1+2\nu \left(\hat{H}_{\text{eff}} [g_{\mu\nu}^{\text{eff}}]-1 \right)}\,,
$
where $\hat{H}_{\text{eff}} [g_{\mu\nu}^{\text{eff}}]$ is an effective Hamiltonian, which will be constructed in the following sections.

\subsection{Action in the EOB theory  for spinning binaries}

It is well established that when a compact binary system emits gravitational waves, it loses energy, thereby producing a gravitational RRF. Consequently, the RRF associated with the ``plus" and ``cross" polarizations of the gravitational waves emitted by the binary can be obtained from the corresponding gravitational-wave energy fluxes. In other words, computing the RRF for these modes requires identifying the null-tetrad component of the perturbed Weyl tensor, $\psi^B_{4}$.

It is essential to note that, within a self-consistent EOB framework, all formulae and quantities including the Hamiltonian, the stress-energy tensor, the gravitational-wave energy flux, and the RRF should be derived from a unified physical model. In this context, we aim to derive expressions for the Hamiltonian and the stress-energy tensor from a single action. Following Ref. \cite{Dixon}, the action in the effective spacetime $g^{\text{eff}}_{\mu\nu}$ can be written as
\begin{align}\label{action}
I = \int d \sigma  \, L\left[x^\mu, v^{\mu}, g^{\text{eff}}_{\mu\nu},  \Omega^{\mu\nu}\right]\,,
\end{align}
where $\sigma $ is a parameter along the representative worldline,  $v^\mu\equiv d x^\mu/d\sigma $ is the tangent vector to the representative worldline, the antisymmetric tensor $\Omega^{\mu\nu}$ characterizes how the tetrad $e^\mu_a$ carried by the effective particle rotates along the worldline. It is defined by 
\begin{equation}\label{omega_def}
\Omega^{\mu\nu}=\eta^{ a b}\, e_{ a}^{\mu}\, \frac{D e_{ b}^{\nu}}{D\sigma }=
{e}^{ a\mu}\,\frac{d{e}_{ a}^{\nu}}{d\sigma } +\Gamma^\mu_{\alpha\beta}\,g_{\text{eff}}^{\alpha\nu}\,v^\beta\,.
\end{equation} 
In the action (\ref{action}), the dynamical variables are the position and the tetrad. The particle's four-momentum $p_{\mu}$ and spin tensor $S_{\mu\nu}$ are defined as  \cite{Dixon}
\begin{align}\label{qsss}
	p_{\mu} \equiv \frac{\partial L}{\partial v^{\mu}} \,,  \ \ \ \  S_{\mu\nu} \equiv 2 \frac{\partial L}{\partial \Omega^{\mu\nu}}, 
\end{align}
where  $p_\mu$ and $S_{\mu\nu} $ satisfy $p^\mu p_\mu=-m_0^2$ and  $ \frac{1}{2}S^{\mu\nu} S_{\mu\nu}=S^2$. Note that $p_\mu$ is not the momentum conjugate to the coordinates $x^\mu$ because $\Omega^{\mu\nu}$ depends on $v^\mu$, as indicated by Eq. (\ref{omega_def}).  We will derive the general formulas for  the Hamiltonian and the stress-energy tensor for gravitational radiation form the action (\ref{action}).

\subsection{Effective Hamiltonian in the EOB theory for spinning binaries}

To construct the effective Hamiltonian, we introduce a  specific 3+1 decomposition of the effective background spacetime. Setting $\sigma=t$ in which $t$ is the time coordinate of that particular decomposition, noting that the parameters $\phi^a$ in the Lorentz transformation  $e_{a}^{\mu}(\phi,x) = \Lambda^{A}_{a} (\phi)\tilde{e}_{A}^{\mu}(x) $ together with their time derivatives, enter the Lagrangian only through the antisymmetric tensor $\Omega^{\mu\nu}$, which is described by $
\Omega^{\mu\nu} = \eta^{a b}\,e^\mu_a (\phi, x)\,\left [\frac{d \phi^a}{d t}\,\frac{\partial e^\nu_{b}}{\partial\phi^a}(\phi, x) + u^\beta\,e^{\nu}_{b,\beta}(\phi, x) \right ]  + \Gamma^\nu_{\alpha\beta}\,g^{\mu\alpha}\,u^\beta\,, $ 
we can rewrite the action (\ref{action}) as  
\begin{align}\label{action0}
I = \int d t  \, L\left[t, x^i, v^{i}, \phi^a,  \dot{\phi}^a\right],
\end{align}
where $v^i=\frac{d x^i}{d \sigma}=\frac{d x^i}{d t}$ and $v^0=\frac{d x^0}{d \sigma}=\frac{d x^0}{d t}=1$. By calculating the momenta $P_i$ and $P_{\phi^a}$ conjugate to $x^i$ and $\phi^a$ from the total variation of the Lagrangian and performing the standard Legendre transformation $H=P_i v^i+P_{\phi^a} \dot{\phi}^a -L$  \cite{BarausseH}, the effective Hamiltonian can be expressed as 
\begin{eqnarray}
H_{\text{eff}}[g_{\mu\nu}^{\text{eff}}] &=& \beta^i\,p_i+\alpha\,\sqrt{m_0^2 +\gamma^{ij}\,p_i\, p_j+ {\cal Q}_4(p)} - E_{t\mu\nu}S^{\mu\nu} \,, \nonumber \\ \label{Hamiltonian1}
\end{eqnarray}
where  $\alpha= \frac{1}{\sqrt{g_{\text{eff}}^{tt}}}\,,$ $ \beta^i = \frac{g_{\text{eff}}^{ti}}{g_{\text{eff}}^{tt}}\,,$ $
\gamma^{ij} = -\Big(g_{\text{eff}}^{ij}-\frac{g_{\text{eff}}^{ti}\,g_{\text{eff}}^{tj}}{g_{\text{eff}}^{tt}}\Big)\,,$  $ E_{\lambda\mu\nu} =  \frac{1}{2}\,\eta_{AB}\,\tilde{e}_\mu^A\,\tilde{e}_{\nu;\lambda}^{B},$  the quantity ${\cal Q}_4(p)$ is a quartic term in the space momenta $p_i$, which was introduced in Ref.~\cite{Damour20002}.

\subsection{Stress-energy tensor for gravitational radiation in the EOB theory}

In this subsection, we derive the stress-energy tensor for  an effective test particle with mass  and spin orbits around a massive spinning black hole. The stress-energy tensor acts as the source of the null tetrad component of the gravitational perturbed Weyl tensor  $\psi^B_{4}$.  

Using the Dirac delta function $\delta^{4}(x-z(\sigma))$, the action in \eqref{action} can be written as 
\be\label{action1}
	I = \int \ud^{4}x \sqrt{-g(x)} \int \ud \sigma \, L\left[x^\mu,  v^{\mu},g^{\text{eff}}_{\mu\nu}, \Omega^{\mu\nu}\right] \frac{\delta^{4}(x-z(\sigma))}{\sqrt{-g(x)}} \,.
\ee
By fixing the worldline and the Lorentz matrices $\Lambda_{A}^{\ph{A}a}$, we define the stress-energy tensor of the particle by taking the variation with respect to the tetrad $\tilde{e}_{A}^{\ph{a}\mu}$, as  \cite{Mino,Dixon1}
$
	T^{\mu\nu}=\frac{1}{\sqrt{-g}} \tilde{e}^{A(\mu}\frac{\delta\, I}{\delta \tilde{e}^{A}_{\ph{a}\nu)}} \,.
$
After carrying out this variation and integrating by parts, the pole-dipole contribution to the stress-energy tensor is obtained as 
\begin{align}
T^{\mu \nu}=\int d\tau\big\{\frac{\delta^{ 4}(x-z(\tau))}{\sqrt{-g}}p^{(\mu}v^{\nu)} -\nabla_{\gamma}\big(S^{\gamma
(\mu}v^{\nu)}\frac{\delta^{4}(x-z(\tau))}{\sqrt{-g}}\big)\big\},
\label{energy-momentum tensor1}
\end{align}
where  $x^\mu=\bigl(t,r(t),\theta(t),\varphi(t)\bigr)$  denotes a geodesic
trajectory, $\tau$ is the proper time along the geodesic,  and 
$v^{\mu}(\tau)=d z^{\mu}(\tau)/d\tau$.  

In the following, we use  Eqs.~(\ref{Hamiltonian1}) and  (\ref{energy-momentum tensor1}) to derive explicit expressions for the effective Hamiltonian and the RRF in the EOB theory. These expressions enter the dynamical evolution equation in Eq.~(\ref{HEq}) within the framework of the effective rotating spacetime.

\section{ Hamiltonian in the EOB theory for a spinning two-body system}\label{Hamiltonian}

In this section, we apply the effective rotating metric defined in Eq. (\ref{effmetric}) to the Hamiltonian in Eq. (\ref{Hamiltonian1}) and derive a general explicit expression for the effective Hamiltonian in the EOB theory. 
 
For the effective rotating metric (\ref{effmetric}), we choose the orthogonal tetrad  $\tilde{e}_\mu^{ A}$ associated with the  background spacetime as 
\begin{align}
\label{OT}
\tilde{e}^{T}_{\mu}   &= \Bigl(\sqrt{{\frac{\Delta}{\Sigma}}}, ~0, ~0, 
-a\sin^2\theta\sqrt{{\frac{\Delta}{\Sigma}}}\Bigr), \  
\tilde{e}^{1}_{\mu}   = \Bigl(0, ~\sqrt{{\frac{\Sigma}{\Delta}}}, ~0, ~0 \Bigr),
\nonumber \\
\tilde{e}^{2}_{\mu}&=  \bigl(0,~0, \sqrt{\Sigma}, ~0 \bigr),
\ 
\tilde{e}^{3}_{\mu}= \bigl(-{\frac{a }{ \sqrt{\Sigma}} }\sin\theta 
,0 ,0 , \frac{r^2+a^2}{  \sqrt{\Sigma}} \sin\theta \bigr).
\end{align}
The choice of this orthogonal tetrad is motivated by the fact that its relation to the null tetrad, introduced in the following section, is simple and transparent. 

Using the relation \(P_i = p_i + E_{i\mu\nu} S^{\mu\nu}\) \cite{BarausseH}, we can rewrite the Hamiltonian in Eq. (\ref{Hamiltonian1}) to linear order in the particle's spin as $
\bar{H}_{\text{eff}} [g_{\mu\nu}^{\text{eff}}]= \bar{H}_{\rm NS} + \bar{H}_{\rm S},
$
 with
$
{\bar{H}}_{\rm NS} = \beta^i \, P_i + \alpha \sqrt{m_0^2 - \gamma^{ij}\,P_i\,P_j + {\cal Q}_4(P)}\,,
$ $ 
{\bar{H}}_{\rm S} = \big(\beta^i\,F_i  + F_t  - \frac{\alpha \gamma^{ij}\,P_i\,F_j }{\sqrt{m_0^2 +
\gamma^{ij}P_iP_j}}\big)\,S_K \,,
$
where $\bar{H}_{\rm NS}$ represents the Hamiltonian for a nonspinning particle in the effective rotating metric \cite{Damour20002}, $S^I=\frac{1}{2}\epsilon^{IJK} S^{\mu\nu} \tilde{e}^J_\mu \tilde{e}^K _\nu$, and
$
F_\mu  = \left(2E_{\mu TI}\,\frac{\bar{\omega}_J}{\bar{\omega}_T} + E_{\mu IJ}\right)\epsilon^{IJK},$
with
 $
\bar{\omega}_\mu=\bar{P}_\mu-m_0 \,\tilde{e}^{T}_\mu,$ $
 \bar{P}_i = P_i,\ $ $
\bar{P}_t = -\beta^i\,P_i-\alpha\, \sqrt{m_0^2 -\gamma^{ij}\,P_i\, P_j},$ $\label{bomegaT}
\bar{\omega}_T = \bar{\omega}_\mu\,\tilde{e}^\mu_{T}=\bar{P}_\mu
 \tilde{e}^\mu_{T}-m_0\,,\ $ and $
\bar{\omega}_I =\bar{\omega}_\mu\,
 \tilde{e}^\mu_{I}= \bar{P}_\mu\,\tilde{e}^\mu_{I}.
$

After a lengthy but straightforward calculation, we find that, in the orthogonal tetrad defined in Eq. (\ref{OT}), the Hamiltonian for the effective rotating spacetime defined by (\ref{effmetric}) can be expressed as
\begin{equation}\label{HH}
\bar{H}_{\text{eff}} [g_{\mu\nu}^{\text{eff}}] = \bar{H}_{\rm NS} + \mathcal{K}^I S_I\,,
\end{equation}
where   $\mathcal{K}^I$ and $S_I$ are described by 
\begin{align}
\mathcal{K}^1 &= -\frac{a\, r\, \Delta\, \sin\theta}{\sqrt{Q\, \Lambda \, \Sigma^3}}\hat{P}_r\frac{\omega_2}{\omega_T} +\frac{a\, \Delta\, \cos\theta}{\sqrt{Q\, \Lambda \, \Sigma^3}}\hat{P}_\theta \frac{\omega_2}{\omega_T} \nonumber \\ &-\Big[\frac{\sqrt{\Delta}\, \cos\theta}{\sin^2\theta \sqrt{Q\, \Lambda \, \Sigma}}-\frac{a\, \Delta\, \sqrt{\Sigma}\,\cot\theta}{ \sqrt{Q\, \Lambda^3}}\frac{\omega_3}{\omega_T} \Big]\hat{P}_\varphi\nonumber 
\\
&+\frac{a \sqrt{\Delta}\,\sin\theta\,\cos\theta\, \omega_{pm}}{  \Lambda\,\Sigma}\frac{\omega_3}{\omega_T},   \nonumber \\
\mathcal{K}^2&= \frac{a\, r\, \Delta\, \sin\theta}{\sqrt{Q\, \Lambda \, \Sigma^3}}\hat{P}_r\frac{\omega_1}{\omega_T} -\frac{a\, \Delta\, \cos\theta}{\sqrt{Q\, \Lambda \, \Sigma^3}}\hat{P}_\theta \frac{\omega_1}{\omega_T} \nonumber \\ &+\Big[\frac{r\, \Delta\,\sqrt{\Sigma}}{\sin\theta \sqrt{Q\, \Lambda^3}}-\frac{a \sqrt{\Delta}\,(\xi +r\,\Sigma)}{ \sqrt{Q\, \Lambda^3\,\Sigma} }\frac{\omega_3}{\omega_T} \Big]\hat{P}_\varphi\nonumber 
\\
&+\Big[\frac{r\, \sqrt{\Delta}\,\sin\theta\, \omega_{pm}}{  \Lambda\,\Sigma}+\frac{(r^2+a^2)\xi -a\,r\,\sin^2\theta\, \omega_{pm}}{  \Lambda \,\Sigma }\frac{\omega_3}{\omega_T}\Big],  \nonumber \\
\mathcal{K}^3 &=- \frac{a^2\, \Delta\, \sin\theta\, \cos\theta}{\sqrt{Q\, \Lambda \, \Sigma^3}}\hat{P}_r -\frac{r\, \Delta}{\sqrt{Q\, \Lambda \, \Sigma^3}}\hat{P}_\theta \nonumber \\ & +\Big[\frac{a\, \sqrt{\Delta}\big((\xi+r\, \Sigma)\,\omega_2-\sqrt{\Delta}\,\Sigma \,\cot\theta\,\omega_1\big)}{\omega_T\,\sqrt{Q\, \Lambda^3\Sigma}}\Big]\hat{P}_\varphi\nonumber 
\\
&-\Big[\frac{a\, \sqrt{\Delta}\,\sin\theta\, \cos\theta\, \omega_{pm}}{  \Lambda\,\Sigma}\frac{\omega_1}{\omega_T}\nonumber \\ &+\frac{\Lambda\,\xi -a\,\sin^2\theta(\xi+r\,\Sigma ) \omega_{pm}}{  \Lambda \,\Sigma^2 }\frac{\omega_1}{\omega_T}\Big], \label{KI2}\\ 
S_1&= \sin\theta \big[a S^{\theta t}+(r^2+a^2)S^{\varphi \theta}\big], \nonumber \\   
S_2&= \frac{\sin\theta}{\sqrt{\Delta}} \big[a S^{t r }+(r^2+a^2)S^{r \varphi}\big],\nonumber \\  
S_3&= \frac{\Sigma}{\sqrt{\Delta}} S^{\theta r},\label{SI}
\end{align}
where $\xi=a^2\,r\,\sin^2\theta-r\, \Delta +\frac{1}{2}\Sigma\,\Delta ',$ \ $ Q = 1 +\frac{\Delta  }{\Sigma} \hat{P}_r^2+\frac{1}{\Sigma} \hat{P}_\theta^2+\frac{\Sigma}{\Lambda \sin^2\theta} \hat{P}_\varphi^2\,,
$
 \ 
 $\omega_T=-\Big(1+\frac{r^2+a^2}{\sqrt{\Lambda}}\sqrt{Q}\Big)+\frac{a \sqrt{\Delta \Sigma}}{\Lambda}\hat{P}_\varphi,
$\ $
\omega_1=\sqrt{\frac{\Delta}{\Sigma}}\hat{P}_r,
$ \ $
\omega_2=\frac{1}{\sqrt{\Sigma}}\hat{P}_\theta,
$ \ $
\omega_3=-a \sin\theta \sqrt{\frac{\Delta}{\Lambda}}\sqrt{Q}+\frac{(r^2+a^2) \sqrt{\Sigma}}{\Lambda \sin\theta}\hat{P}_\varphi,$
\  $\hat{P}_i \equiv P_i/m_0$,  and  $S^{\mu\nu}=\epsilon^{\mu\nu\alpha\beta}u_\alpha S_\beta$, with $\epsilon^{\mu\nu\alpha\beta}$  is the four-dimensional antisymmetric Levi-Civita symbol. Here and hereafter, a prime denotes derivation with respect to $r$.

\section{Null tetrad component of perturbed  gravitational Weyl tensor \texorpdfstring{$\psi^B_{4}$}{}} 

 It is well known that, in order to determine the radiation reaction force of the ``plus'' and ``cross'' modes of GWs emitted by spinning binaries, one must identify the null-tetrad component of the perturbed gravitational Weyl tensor, \( \psi^B_{4} = \frac{1}{2}(\ddot h_{+} - i\ddot h_{\times}) \), including source terms in the effective rotating spacetime. In this section, we derive a decoupled, variable-separated Teukolsky-like equation for $\psi^B_4$ with a source term and construct a formal solution to this equation.
 
 \subsection{Decoupled and separable equation for \texorpdfstring{ $\psi^B_{4}$}{} in the effective rotating spacetime}

In the effective rotating spacetime described by Eq. (\ref{effmetric}), we choose 
\begin{align}\label{NTetrad}
 l^{\mu}&=\sqrt{\frac{\Sigma}{\Delta}}(\tilde{e}^\mu_T+\tilde{e}^\mu_1), \ \ \ \   \ \ \ 
 n^{\mu}=\frac{1}{2}\sqrt{\frac{\Delta}{\Sigma}}(\tilde{e}^\mu_T-\tilde{e}^\mu_1),   \nonumber \\
 m^{\mu}&=\frac{1}{\bar{\rho}}\sqrt{\frac{\Sigma}{2}}(\tilde{e}^\mu_2+i \tilde{e}^\mu_3),  \ \ \ \   \bar{m}^{\mu}=m^{\mu*}.
\end{align}
Using the Newman-Penrose formalism to describe the gravitational perturbation $g_{\mu \nu}=g_{\mu \nu}^{\text{eff}}+\varepsilon h_{\mu \nu}^{B}$, we  obtain \cite{Jing2025SC} 
\begin{align}
&\Big[({\bf{\Delta}}+3\gamma-\bar{\gamma}+4\mu+\bar{\mu})(D+4\epsilon-\rho)-(\bar{\delta}+3\alpha+\bar{\beta}-\bar{\tau}+4\pi)\nonumber \\ & \times (\delta+4\beta-\tau)  -3\Psi_{2} -12 f \Lambda \Big]\Psi_{4}^{B}
=T^B_{4} +{\cal{G}}_4^B,
\label{eq06J1}
\end{align}
with
\begin{align}
T^B_{4}&=({\bf{\Delta}}+3\gamma-\bar{\gamma}+4\mu+\bar{\mu})
\big[(\bar{\delta}-2\bar{\tau}+2\alpha)\Phi_{21}^{B}
-({\bf{\Delta}}+2\gamma\notag\\
& -2\bar{\gamma}+\bar{\mu})\Phi_{20}^{B}\big]
-(\bar{\delta}+3\alpha+\bar{\beta}-\bar{\tau}+4\pi)
\big[(\bar{\delta}-\bar{\tau}+2\bar{\beta}\notag\\
& +2\alpha)\Phi_{22}^{B}
-({\bf{\Delta}}+2\gamma+2\bar{\mu})\Phi_{21}^{B}\big],\\
 {\cal{G}}_4^B&=\big[({\bf{\Delta}}+3\gamma-\bar{\gamma}+4\mu+\bar{\mu})(\bar{\delta}+2\alpha+4\pi)
-(\bar{\delta}+3\alpha+\bar{\beta}-\bar{\tau}\notag\\
& +4\pi)({\bf{\Delta}}+2\gamma+4\mu)\big] \psi_{3}^{B}
-4\big[({\bf{\Delta}}+3\gamma-\bar{\gamma}+2\mu+2\bar{\mu})\notag\\
& \times (\Phi_{11}\lambda^{B})+\nu^{B}(\bar{\delta}+\pi-\bar{\tau})\Phi_{11}+\frac{1}{2} (\Phi_{11}+6 f \Lambda)\Psi_{4}^{B}\big],\end{align}
where  $ \Psi_{2}=\frac{1}{12 \bar{\rho} \bar{\rho}^{*3} }\big[12\Delta-6 \bar{\rho}^{*}\Delta^{\prime} + \bar{\rho}^{*2} \Delta^{\prime\prime}-2\big(\bar{\rho}^{*2} +6 a (a+ i r \cos\theta)\big)\big]$, and $ \phi_{11}=\frac{1}{8 \bar{\rho}^2 \bar{\rho}^{*2} }\big[4\Delta-4 r \Delta^{\prime} + \bar{\rho}\bar{\rho}^{*} \Delta^{\prime\prime}+2 r^2 -5 a^2 - a^2 \cos(2\theta)\big]$, $\Lambda=-\frac{R}{24}=\frac{2-\Delta''}{24 \bar{\rho}\bar{\rho}^{*}}$ ( where $R$ is the Ricci scalar curvature), and the factor $f$ should be fixed in the specific physical process. 

It is essential to note that the system contains four unknown functions, $\nu^{B}$, $\lambda^{B}$, $\Psi_{3}^{B}$, and $\Psi_{4}^{B}$, but only three equations  \cite{Jing2025SC}. Therefore, we may set
$
{\cal{G}}_4^B=0. \label{gauge}
$
By substituting $\Psi_{4}^{B}=(\bar\rho^{*})^{-4} \phi_{4}^{B}$ and $T_4^B \equiv  {\cal T}^{(-2)}/(2 \rho\rho^{*5} )$, 
and taking  \( \phi_{4}^{B} \) and \( {\cal T}^{(-2)} \)  as 
 \begin{align}
&\phi_{4}^{B}=\int d\omega \sum_{l,m} R^{(-2)}_{lm\omega}(r){}^{}_{s}S^{a\omega}_{lm}(\theta)e^{-i\omega t}e^{im\varphi},
\\
&{\cal T}^{(-2)}=\int d\omega \sum_{l,m} {\cal T}^{(-2)}_{lm\omega}(r){}^{}_{s}S^{a\omega}_{lm}(\theta)e^{-i\omega t}e^{im\varphi},\label{TTT}
\end{align}
we obtain the separated equations 
 \begin{align}
&\Delta^{2}\frac{d}{dr}\Big(\frac{1}{\Delta}\frac{d R^{(-2)}_{lm\omega}}{dr}\Big)
+\Big[\frac{K^{2}+2 i  K\Delta'}{\Delta}+\Big(f-\frac{1}{2}\Big) (2-\Delta'')\nonumber \\ & \ \ -8i  \omega r-\boldsymbol{\lambda}\Big]R^{(-2)}_{lm\omega}={\cal T}^{(-2)}_{lm\omega},
\label{defTE1}
\\
&\frac{1}{\sin\theta}\frac{d}{d\theta}\Big(\sin\theta\frac{d {}_{-2}S^{a\omega}_{lm}}{d\theta}\Big)
+\Big(a^{2}\omega^{2}\cos^{2}\theta-\frac{m^{2}-4  m\cos\theta}{\sin^{2}\theta}\nonumber \\ & \ \  + 4 a\omega\cos\theta  -4\cot^2\theta
-2+(l+2)(l-1)\Big) {}_{-2}S^{a\omega}_{lm}=0,
\label{swsh}
\end{align}
where  $\boldsymbol{\lambda}=(l+2)(l-1)+a^2\omega^2-2 a m \omega$. 
Eq. (\ref{defTE1}) reduces to the radial Teukolsky equation in Kerr spacetime for the special case in which $a_2=a_3=a_4=0$.

\subsection{ Source  of the gravitational radiation}

Introducing the tetrad basis \( Z_{a\mu}=(l_\mu,\, n_\mu,\, m_\mu,\, \bar{m}_\mu) \) for the effective background spacetime, where \( l_\mu, n_\mu, m_\mu, \) and \( \bar{m}_\mu \) are defined by Eq. (\ref{NTetrad}), the projections of the energy-momentum tensor in Eq. (\ref{energy-momentum tensor1}), onto the null tetrad are given by 
\begin{align}
T_{a b}&=Z_{a\mu}Z_{b\nu}T^{\mu\nu}(x)\nonumber \\ &=\int d\tau\Big\{\frac{\delta^{(4)}(x-z(\tau))}{\sqrt{-g}}\Big[ m_0 u_{(a}v_{b)}-\eta^{pi}\eta^{q j}\Big(S_{ij}v_{(a} \gamma_{b)q p}\nonumber \\ & \ \  +S_{i(a} \gamma_{b)q p}v_j\Big)\Big]  -\frac{1}{\sqrt{-g}}\frac{\partial }{\partial x^{\gamma}}\Big(S^{\gamma}_{(a}
v_{b)}\delta^{(4)}(x-z(\tau))\Big)\Big\}.
\label{energy-momentum tensor Null}
\end{align}
Equation (\ref{energy-momentum tensor Null}) implies that the tetrad components of the energy-momentum tensor are 
\begin{align}
&T_{n n}=   \frac{1 }{ \sin\theta}\Big[ B_{nn}\delta(r-r(t))-\frac{\partial}{\partial r}(D_{nn}\delta(r-r(t)))\Big]\nonumber \\ & \ \ \ \ \ \times \delta(\theta-\theta(t)) \delta(\varphi-\varphi(t)),
\nonumber\\
&T_{{\bar m} {\bar m}}= \frac{1 }{ \sin\theta} \Big[B_{\bar m\bar m}\delta(r-r(t))-\frac{\partial}{\partial r}(D_{\bar m\bar m} \delta(r-r(t)))\Big]\nonumber \\ & \ \ \ \ \ \times \delta(\theta-\theta(t)) \delta(\varphi-
\varphi(t)),\nonumber\\
&T_{n {\bar m} }=   \frac{ 1 }{ \sin\theta}\Big[B_{n\bar m}\delta(r-r(t))-\frac{\partial}{\partial r}(D_{n \bar m}\delta(r-r(t)))\Big]\nonumber \\ & \ \ \ \ \ \times   \delta(\theta-\theta(t)) \delta(\varphi-\varphi(t)),
 \label{tij}
\end{align}
where \(B_{nn}\), \(B_{\bar{m}\bar{m}}\), \(B_{n\bar{m}}\), \(D_{nn}\), \(D_{\bar{m}\bar{m}}\), and \(D_{n\bar{m}}\) are 
\begin{align}
B_{n n}&=\frac{1}{ \Sigma \dot{t} }\Big\{m_0 u_n v_n-2\Big[(\gamma+\bar{\gamma})S_{ln}+(\alpha+\bar{\beta})S_{n m}+(\bar{\alpha}+\beta)S_{n\bar{m}}\nonumber \\ & -\frac{1}{2}(\pi S_{n m} +\bar{\pi}S_{n\bar{m}})\Big]v_n+\mu S_{n\bar{m}}v_m+\bar{\mu}S_{nm}v_{\bar{m}}-i (\omega S^t_n\nonumber \\ & -m S^\varphi _n)v_n+D_{nn}(\Sigma \dot{t} )'\Big\},
\nonumber\\ 
B_{\bar m \bar m}&=\frac{1}{ \Sigma \dot{t} }\Big\{ m_0 u_{\bar m}v_{\bar m}-\Big[\bar \tau S_{l \bar{m}}v_n-\pi S_{n \bar{m}}v_l+\Big((\pi+\bar{\tau})S_{l n}+(\rho^* \nonumber \\ & +2\epsilon-2\bar{\epsilon})S_{n\bar{m}}+( 2\gamma-2\bar{\gamma}-\mu)S_{l\bar m}\Big)v_{\bar m}\Big] -i (\omega S^t_{\bar m}\nonumber \\ & -m S^\varphi _{\bar m})v_{\bar m}+D_{\bar m \bar m}(\Sigma \dot{t} )'\Big\},\nonumber\\ 
B_{n \bar m}&=\frac{1}{ 2 \Sigma \dot{t} }\Big\{ m_0 (u_{n}v_{\bar m}+u_{\bar m}v_{n})-\Big[\Big((\pi+ 2 \bar{\tau})S_{ln} +(2 \gamma-\mu) S_{l \bar{m}}\nonumber \\ & +2 (\rho^* +\epsilon)S_{n \bar{m}}\Big) v_n-\mu S_{n\bar{m}}v_l-\pi S_{n \bar{m}}v_m+\Big(2\gamma S_{l n}+(- \pi \nonumber \\ & +2\alpha)S_{n m}+2(\beta-\bar{\pi})S_{n \bar{m}}\Big)v_{\bar m}\Big] -i \omega( S^t_{\bar m}v_n+S^t_nv_{\bar m})\nonumber \\ & +i m (S^\varphi _{\bar m} v_{n}+S^\varphi_{n}v_{\bar m}) +D_{n \bar m}(\Sigma \dot{t} )'\Big\}, \\
D_{n n}&=\frac{S^r_n v_n}{ \Sigma \dot{t} }, \ \ 
D_{\bar m \bar m}= \frac{S^r_{\bar m} v_{\bar m}}{\Sigma \dot{t}}, \ \  
 D_{n \bar m}=
\frac{(S^r_{\bar m} v_{n}+S^r_{n} v_{\bar m})}{ 2 \Sigma \dot{t}}, \label{Ncij}
\end{align}
in which $\dot t=dt/d\tau$,  and 
\begin{align}
& \rho=-\frac{1}{\overline{\rho}^{*}}, \ \  \mu=-\frac{\Delta }{2\overline{\rho}\overline{\rho}^{*2} },  \ \ \gamma=\frac{\Delta ^{\prime}}{4\overline{\rho}\overline{\rho}^{*}}+\mu,  \ \ \pi=\frac{i a\sin\theta}{\sqrt{2}\overline{\rho}^{*2} }, \nonumber \\ &  \tau=-\frac{i a \sin\theta}{\sqrt{2}\overline{\rho}\overline{\rho}^{*}},\ \  \beta=\frac{\cot\theta}{2\sqrt{2}\overline{\rho}}, \ \  \alpha=\pi-\beta^* ,   \nonumber   \\ 
&v_l=l_\mu v^{\mu}, \quad  v_n=n_\mu v^{\mu},\quad v_{ m}=m_\mu v^{\mu}, \quad   v_{\bar m}=\bar{m}_\mu v^{\mu},  \nonumber\\ 
&S_{ln}=l_\mu n_\nu S^{\mu\nu}, \    S_{l m}=l_\mu m_\nu S^{\mu\nu},\ \   S_{l\bar{m}}=S_{l m}^{*},\nonumber \\ &   S_{n m}=n_\mu m_\nu S^{\mu\nu},  \ 
S_{n \bar{m}}=S_{n m}^{*},  \   S^t_n= n_\nu S^{t\nu}, \ 
S^r_n= n_\nu S^{r\nu}, \nonumber \\ &   S^{\varphi}_n= n_\nu S^{\varphi \nu},  \  S^t_{\bar m}= {\bar m}_\nu S^{t\nu}, \   S^r_{\bar m}={\bar m}_\nu S^{r\nu}, \  S^{\varphi}_{\bar m}={\bar m}_\nu S^{\varphi \nu},
\nonumber \\   
 &
 S^t_l= l_\nu S^{t\nu}, \
S^r_l= l_\nu S^{r\nu}, \    S^{\varphi}_l= l_\nu S^{\varphi \nu},  \  S^t_{ m}= { m}_\nu S^{t\nu}, \nonumber \\ &   S^r_{m}={ m}_\nu S^{r\nu}, \  S^{\varphi}_{ m}={ m}_\nu S^{\varphi \nu}.
 \end{align}

 For a source confined to a finite interval in $r$, Eqs. (\ref{TTT}) and (\ref{tij}) yield \begin{align}
{\cal T}^{(-2)}_{\ell m \omega}&=m_0 G \int^{\infty}_{-\infty}dt 
e^{i\omega t-i m \varphi(t)}
\Delta  ^2\Big\{\big(A_{nn\,0} B_{nn}+A_{{\bar m}n\,0} B_{{\bar m}n}\nonumber \\ & +
A_{{\bar m}{\bar m}\,0} B_{{\bar m}{\bar m}}\big) \delta(r-r(t)) +\big[\big(A_{{\bar m}n\,1} B_{{\bar m}n}+A_{{\bar m}{\bar m}\,1} B_{{\bar m}{\bar m}} \big)\nonumber \\ & \times \delta(r-r(t))\big]
' +\big(A_{{\bar m}{\bar m}\,2}B_{{\bar m}{\bar m}}\delta(r-r(t))\big)
'' 
\nonumber \\ &
-\Big[\big(A_{nn\,0} (D_{nn}\delta(r-r(t)))'+A_{{\bar m}n\,0} (D_{{\bar m}n}\delta(r-r(t)))'\nonumber \\ & +
A_{{\bar m}{\bar m}\,0} (D_{{\bar m}{\bar m}}\delta(r-r(t)))'\big)  +\Big(A_{{\bar m}n\,1} (D_{{\bar m}n}\delta(r-r(t)))'\nonumber \\ & +A_{{\bar m}{\bar m}\,1} (D_{{\bar m}{\bar m}}\delta(r-r(t)))' \Big)
' \nonumber \\&
+\Big(A_{{\bar m}{\bar m}\,2}(D_{{\bar m}{\bar m}}\delta(r-r(t)))'\Big)
''\Big]\Big\}_{\theta(t)},
\label{NTgenTTsl}
\end{align}
where  
\begin{align}
A _{nn\,0}&=-\frac{2 \bar\rho (\bar\rho^{*})^2}{ \sqrt{2\pi}\,\Delta ^2 }\,
\,
\mathscr{L}_1^{\dag}\Big[(\bar\rho^{*})^4\mathscr{L}_2^{\dag}\Big(\frac{1}{(\bar\rho^{*})^3}\,_{-2}S_{\ell m}(\theta)\Big)  \Big],\nonumber \\
A _{{\bar m}n\,0}&=\frac{2 }{ \sqrt{\pi}\Delta   } 
\,   (\bar\rho^{*})^3 \Big[\Big(\frac{i K}{ \Delta   }+\frac{1}{\bar\rho}+\frac{1}{\bar\rho^{*}}\Big) \mathscr{L}_2^{\dag}\nonumber \\ & -a \sin\theta \frac{K}{\Delta  }\Big(\frac{1}{\bar\rho}-\frac{1}{\bar\rho^{*}}\Big) \Big]\,_{-2}S_{\ell m}(\theta) ,\nonumber \\
A _{{\bar m}{\bar m}\,0}
&=\frac{1 }{ \sqrt{2\pi}}\frac{(\bar\rho^{*})^3}{\bar\rho}\,  
 \Bigl[
i\Bigl(\frac{K }{ \Delta   }\Bigr)' +\frac{(K)^2 }{ (\Delta   )^2}\nonumber - 2 i\frac{ K }{ \Delta   \bar\rho^{*}}\Bigr]\,_{-2}S_{\ell m}(\theta) ,\nonumber \\
A _{{\bar m}n\,1}&=\frac{
 2 (\bar\rho^{*})^3}{ \sqrt{\pi}\Delta    }\,
 \Big[ \mathscr{L}_2^{\dag}+i a \sin\theta\Big(\frac{1}{\bar\rho}-\frac{1}{\bar\rho^{*}}\Big) \Big]\,_{-2}S_{\ell m}(\theta) 
,\nonumber \\
A _{{\bar m}{\bar m}\,1}
&=-\frac{ 2  (\bar\rho^{*})^3 }{ \sqrt{2\pi} \bar\rho}
\,
\Bigl(  \frac{i K}{ \Delta   }+\frac{1}{\bar\rho^{*}}\Bigr)\,_{-2}S_{\ell m}(\theta)  ,\nonumber \\
A _{{\bar m}{\bar m}\,2}
&=-\frac{(\bar\rho^{*})^3}{ \sqrt{2\pi \bar\rho}}\,
\,_{-2}S_{\ell m}(\theta).
\label{Aijsk} 
\end{align}

\subsection{ Formal solution of Teukolsky-like equation for \texorpdfstring{ $\psi^B_{4}$}{}  }

With Eq.~(\ref{NTgenTTsl}) at hand, we can solve the radial Teukolsky-like equation (\ref{defTE1}) using the Green's-function method. It is worth noting that, apart from the factor $\tilde Z^{I(-2)}_{\ell m\omega}$, the inhomogeneous solution for $R^{(-2)}_{lm\omega}$ in Eq.~(\ref{defTE1}) at infinity is identical to that in the spinless case \cite{Jing2025SC}. The quantity $\tilde Z^{I(-2)}_{\ell m\omega}$ for spinning binary systems is 
\begin{align}
\tilde Z^{I (-2)}  _{\ell m\omega}&=
\frac{m_0 G}{2i\omega B^{\rm inc}_{\ell m\omega}}
\int^{\infty}_{-\infty}dt e^{i\omega t-i m \varphi(t)}
\Big\{ A^{(-2)}_0
R^{\rm in(-2)}_{\ell m\omega}\nonumber \\ & - A^{(-2)}_1
\Big(R^{\rm in(-2)}_{\ell m\omega}\Big)' + A^{(-2)}_2 \Big(R^{\rm in(-2)}_{\ell m\omega}\Big)'' 
+\Big(A_{nn\,0}R^{\rm in(-2)}_{\ell m\omega}\Big)' \nonumber \\ & \times  D_{nn}  +\Big(A_{{\bar m}n\,0} R^{\rm in(-2)}_{\ell m\omega}\Big)'D_{{\bar m}n}+
\Big(A_{{\bar m}{\bar m}\,0} R^{\rm in(-2)}_{\ell m\omega}\Big)'D_{{\bar m}{\bar m}} \nonumber\\
& 
-\Big(A_{{\bar m}n\,1} \big(R^{\rm in(-2)}_{\ell m\omega}\big)'\Big)' D_{{\bar m}n}-\Big(A_{{\bar m}{\bar m}\,1} \big(R^{\rm in(-2)}_{\ell m\omega}\big)'\Big)' D_{{\bar m}{\bar m}}
\nonumber\\
&
+\Big(\big(R^{\rm in(-2)}_{\ell m\omega}\big)'' A_{{\bar m}{\bar m}\,2}\Big)' D_{{\bar m}{\bar m}}
\Big\}_{r=r(t),\theta= \theta(t)},
\label{ZZSch}
\end{align}
where $A^{(-2)}_0 =\big(A_{nn\,0} B_{nn}+A_{{\bar m}n\,0} B_{{\bar m}n}+
A_{{\bar m}{\bar m}\,0} B_{{\bar m}{\bar m}}\big),$ $
 A^{(-2)}_1 =\big(A_{{\bar m}n\,1} B_{{\bar m}n}+A_{{\bar m}{\bar m}\,1} B_{{\bar m}{\bar m}} \big),$ $
 A^{(-2)}_2 =A _{{\overline m}{\overline m}\,2}B_{{\overline m}{\overline m}}$.
 
For a circular orbit, we set  $r(t) = r_0$,  $\theta(t) = \theta_0$, and $\varphi(t) = \Omega t$, where $ \Omega$ is the orbital angular velocity. Consequently, we obtain the following formal solution of $ \psi_4^B $
\begin{align}
\psi^{B(I)}_4&=\frac{1}{ r}\sum_{\ell m n}
\frac{\pi m_0 G}{i\omega_n B^{\rm inc}_{\ell m\omega_n}}
\Big\{ A^{(-2)}_0
R^{\rm in(-2)}_{\ell m\omega}- A^{(-2)}_1
\big(R^{\rm in(-2)}_{\ell m\omega}\big)' \nonumber \\ & + A^{(-2)}_2 \big(R^{\rm in(-2)}_{\ell m\omega}\big)'' 
+\big(A_{nn\,0}R^{\rm in(-2)}_{\ell m\omega}\big)' D_{nn}\nonumber \\ & +\big(A_{{\bar m}n\,0} R^{\rm in(-2)}_{\ell m\omega}\big)'D_{{\bar m}n}+
\big(A_{{\bar m}{\bar m}\,0} R^{\rm in(-2)}_{\ell m\omega}\big)'D_{{\bar m}{\bar m}} \nonumber\\
&
-\Big[\Big(A_{{\bar m}n\,1} \big(R^{\rm in(-2)}_{\ell m\omega}\big)'\Big)' D_{{\bar m}n}+\Big(A_{{\bar m}{\bar m}\,1} \big(R^{\rm in(-2)}_{\ell m\omega}\big)'\Big)' D_{{\bar m}{\bar m}} \Big]
 \nonumber \\&
+\Big(\big(R^{\rm in(-2)}_{\ell m\omega}\big)'' A_{{\bar m}{\bar m}\,2}\Big)' D_{{\bar m}{\bar m}}
\Big\}_{r_0,\theta_0}\frac{{}_{-2}S_{\ell m} }{ \sqrt{2\pi}}
e^{i\omega_n(r^*-t)},  \nonumber \\ & \ \    \text{for}\  (r \to \infty). \label{psi411}
\end{align}

\section{Energy flux, RRF and waveform for ``plus" and ``cross" modes of GW generated by  spinning binaries}\label{RRFWF}

Note that the null tetrad component of the perturbed Weyl tensor for gravitational  waves is given asymptotically by \(\psi^{B(I)}_4 = \frac{1}{2}(\ddot h_{+} - i \ddot h_{\times})\). Consequently, the energy flux \cite{Ref:poisson,TagoshiSasaki745, WOS:001272341400001,WOS:001753443600001} can be written as
\begin{eqnarray}\label{de}
\frac{dE}{dt} & = &\lim_{r\rightarrow\infty}\Big[\frac{r^2}{4\pi G \omega^2}\int_\theta\int_\varphi\sin\theta\;d\theta\;d\varphi\left| \psi^{B(I)}_4\right|^2\Big] .
\end{eqnarray}

Using Eq.  (\ref{de}), the  reduced RRF for the ``plus" and ``cross" modes of gravitational wave generated by spinning binaries is 
\begin{align}
\hat{\bm{\mathcal{F}}}=\frac{1}{\nu  M  \Omega}\lim_{r\rightarrow\infty}\Big[\frac{r^2}{4\pi G \omega^2}\int_\theta\int_\varphi\sin\theta\;d\theta\;d\varphi\left| \psi^{B(I)}_4 \right|^2\Big] \frac{\vP}{|\vvr\times \vP|}.\nonumber \\ 
 \label{FFdE1}
\end{align}

By comparing $\psi^B_4$ with the waveform \cite{Kidder}
$
h_{+}-i h_{\times}=\sum_{l=2}^{\infty} \sum_{m=-l}^{l} h^{lm}\frac{ \  _{-2}S^{lm}(\theta, \varphi)}{\sqrt{2\pi}} ,
$
we find
\begin{align}
h^{l m}&=
\frac{2 r \pi m_0 G}{i\omega_n^3 B^{\rm inc}_{\ell m\omega_n}}
\big\{ A^{(-2)}_0
R^{\rm in(-2)}_{\ell m\omega}- A^{(-2)}_1
(R^{\rm in(-2)}_{\ell m\omega})' + A^{(-2)}_2 (R^{\rm in(-2)}_{\ell m\omega})'' \nonumber \\ &
+(A_{nn\,0}R^{\rm in(-2)}_{\ell m\omega})' D_{nn}+(A_{{\bar m}n\,0} R^{\rm in(-2)}_{\ell m\omega})'D_{{\bar m}n}+
(A_{{\bar m}{\bar m}\,0} R^{\rm in(-2)}_{\ell m\omega})' \nonumber\\
&
 \times D_{{\bar m}{\bar m}} -\big[(A_{{\bar m}n\,1} (R^{\rm in(-2)}_{\ell m\omega})')' D_{{\bar m}n}+(A_{{\bar m}{\bar m}\,1} (R^{\rm in(-2)}_{\ell m\omega})')' D_{{\bar m}{\bar m}} \big]
 \nonumber \\&
+\big((R^{\rm in(-2)}_{\ell m\omega})'' A_{{\bar m}{\bar m}\,2}\big)' D_{{\bar m}{\bar m}}
\big\}_{r_0,\theta_0} e^{i\omega_n(r^*-t)}.
\label{hform}
\end{align}

Note that all formulas and quantities within the EOB framework are derived from a unified physical model. Therefore, the EOB theory for spinning binaries can be regarded as a self-consistent theoretical framework.

\section{ Conclusions and discussions}
In the context of spinning black-hole binaries, the EOB theory maps the conservative dynamics of two compact objects with masses $m_1$ and $m_2$ and spins $\vS_1$ and $\vS_2$ onto the dynamics of an effective particle with mass $m_0$ and spin $\vS$, orbiting a massive black hole characterized by mass $M$ and spin $\vS_{\text{metric}}$. The dynamical evolution of this system is governed by Hamilton's equations. To construct a self-consistent EOB theory, we systematically derive the required formulas and quantities in the effective spacetime.

Using two types of tetrads, an orthonormal tetrad $e_a^{\mu}$ associated with the worldline of the effective particle, and another tetrad $\tilde{e}^{\mu}_{A}$ associated with the effective background spacetime, we derive the Hamiltonian in Eq. (\ref{Hamiltonian1}) and the stress-energy tensor in Eq. (\ref{energy-momentum tensor1}) at linear order in the particle's spin, starting from the action in Eq. (\ref{action}) for an effective particle with mass and spin orbiting a massive rotating black hole described by the effective rotating metric. The stress-energy tensor serves as the source of the Teukolsky-like equation governing the null-tetrad component $\psi^B_4$ of the perturbed Weyl tensor in the effective background spacetime.

Next, by substituting the effective rotating metric into the Hamiltonian, we obtain the general explicit expressions for the effective Hamiltonian of a spinning two-body system in the EOB theory, as given in Eqs. (\ref{HH}), (\ref{KI2}), and (\ref{SI}).

To determine the RRF associated with the ``plus" and ``cross" gravitational-wave modes generated by spinning binaries, it is necessary to compute $\psi^{B(I)}_{4} = \frac{1}{2}(\ddot h{+} - i\ddot h_{\times})$ with source terms in the effective rotating spacetime. The reduced RRF, denoted by $\hat{\bm{\mathcal{F}}}=\frac{1}{\nu M_0 \Omega |\vvr\times \vP|}\frac{dE}{dt}\vP$ \cite{Buonanno2006}, is related to the gravitational-wave energy flux, $\frac{dE}{dt} = \frac{1}{16\pi G} \int (\dot{h}{+}^2 + \dot{h}{\times}^2) r^2 d\Omega$. Using the separated equations (\ref{defTE1}) and (\ref{swsh}) for the null-tetrad component $\psi^B_{4}$ in the effective rotating spacetime, together with the gravitational-radiation source terms in Eq. (\ref{NTgenTTsl}), we derive the formal solution of the Teukolsky-like equation in Eq. (\ref{psi411}).

Finally, based on the solution of the Teukolsky-like equation, we obtain the RRF in Eq. (\ref{FFdE1}) and the waveform in Eq. (\ref{hform}) for the ``plus" and ``cross" gravitational-wave modes generated by spinning binaries.

A self-consistent EOB theory with 5PM precision is developed to describe the dynamical evolution of spinning binary systems, with all formulas and quantities in the Hamiltonian equations derived within a unified physical framework. 

To validate the theory, we perform numerical calculations using the EOB theory and compare the results with numerical relativity (NR) data from the SXS catalog. At 5PM accuracy, the binding energy-angular momentum relation agrees with NR results up to the innermost stable circular orbit within $4$\textperthousand, $0.8$\textperthousand,  and $0.6$\textperthousand\  for  mass ratio $q = 1, 10$, and $20$, respectively.

The self-consistent EOB framework presented here imposes no restriction on the mass ratio, making it applicable to systems with arbitrary mass ratios, including comparable, intermediate, and extreme mass-ratio cases. It thus bridges the gap between comparable-mass and extreme-mass-ratio binaries.

\section*{Acknowledgement}
{
We would like to thank professors S. Chen,  Q. Pan and X. He for useful discussions on the manuscript. This work was supported by the Grant of NSFC Nos. 12035005 and 12475051, and National Key Research and Development  Program of China No. 2020YFC2201400.
\\
\\
The authors declare that they have no conflict of interest. 
}  

\bibliography{mybib}

\end{document}